\documentclass[epj]{svjour}

\usepackage{graphicx}
\usepackage{fancyhdr}

\def\mytitle{My title}
\def\myauthors{My name}
\def\mytype{My type of session}
\def\mysession{My session}

\def\mytitle{Intersecting Brane Models and Cosmology} 
\def\myauthors{Jason Kumar}    
\def\mytype{Contributed Talk}
\def\mysession{Theoretical Models}

\begin{document}
\title{Intersecting Brane Models and Cosmology}
\author{Jason Kumar\inst{1}
\thanks{\emph{Email:} kumarj@uci.edu}%
}                     
%
%
\institute{Department of Physics, University of California,
Irvine, CA  92697
}
%
\date{}
\abstract{
We discuss the application of general features of intersecting
brane model constructions in cosmology.  In particular,
we describe a scenario for $D$-term inflation which
arises straightforwardly in IBM constructions wherein
open non-vectorlike strings play the role of the inflaton.
We also show that baryogenesis driven by hidden sector
mixed anomalies can naturally arise in intersecting
brane models.
\PACS{
      {PACS-11.25.Uv   }{D branes}   \and
      {PACS-98.80.Cq}{Particle/field-theory models of the early Universe }
     } 
} 
\maketitle
\section{Introduction}
There have been several attempts to use
string theory to make connections between quantum gravity and
the physics which can be observed at low-energies.  Cosmology
has been one particular focus of attention \cite{reviewinflation}.
In recent
years, intersecting brane models (IBMs) in Type II string theory
have been one well-studied scenario\cite{IBM}.  In this
scenario, gravitational
physics is generated as usual by closed strings, while the open
strings stretching between D-branes generate the gauge and matter
degrees of freedom of the Standard Model.

Several explicit intersecting brane models have been constructed,
but it seems clear from the nature of the constructions that
known examples are merely the simplest.  Moreover,
this very simplicity has been somewhat counterproductive, as each
of the known models fails in some way to match low-energy observations.
Indeed, the very large number of possible IBMs suggests that any particular
explicit construction likely does not describe the real world in
detail.  As such, any lesson which is particular to a specific construction
may be irrelevant to more realistic models.

In this work we emphasize a different approach, which is to focus
on general new features which are common to a wide class of IBMs.
The advantage of such an approach
is that, although new physics may be illustrated in a specific simple
example, the results can be exported to more complicated models which
are also more realistic.

In section 2 we describe the general features of intersecting brane
models which are relevant for the new physics which we discuss.
In section 3 we describe a scenario of D-term inflation which
arises naturally in this IBM setup.  In section 4 we describe
a mechanism for baryogenesis which also arises naturally in IBMs, and
we conclude with a discussion of future prospects in section 5.

\section{Intersecting Brane Models}
Type II string theory naturally lives in 10 dimensions.  In an IBM,
one compactifies this space on 6D orientifolded CY 3-fold.  This
reduces the supersymmetry of the theory from 32 real supersymmetries
to the more realistic $N=1$ SUSY in 4D.  The spacetime-filling
orientifold planes introduced by the compactification are charged;
Gauss' Law (equivalently, the RR-tadpole constraints) requires that
this charge be cancelled, and one can do so by introducing D6-branes
(in Type IIA) which fill spacetime and wrap
appropriate 3-cycles of the compactification
manifold.  These branes provide an extra feature - the open strings
which begin and end on these branes generate a gauge theory, as well
as chiral matter.  For a realistic IBM, we would like one sector of
this gauge theory to be SM-like.  But generally we will have additional
sectors, since there is no reason for the branes which provide the
SM gauge theory alone to be precisely sufficient to cancel all tadpoles.

In these models, the gauge bosons arise from strings beginning and
ending on the same brane, and the gauge group is determined by the
number of branes which are stacked on top of each other ($U(N)$ for
branes which do not lie on orientifold planes).  Chiral matter
arises from open strings which stretch from one brane stack to another
(or its orientifold image), and transforms under the bifundamental
representation of the two gauge groups associated with the two brane
stacks.  The number of such multiplets in each representation is
counted by the topological intersection number of the brane stacks.

It is important to note that any two brane stacks will generally have
a non-zero intersection number.  This is true because any two
3-cycles will generically intersect on a 6-manifold.  This implies
that for two gauge groups $U(N_a)$ and $U(N_b)$, there generically
exists non-vectorlike matter transforming in the bifundamental
representation.  This in turn will contribute to a mixed anomaly
of the form $[U(1)_a U(N_b)^2]$, where $U(1)_a$ is the diagonal
subgroup of $U(N_a)$.  Note that although we have specifically
focussed on IBMs in the context of Type IIA string theory, similar
statements follow for T-dual models in Type IIB.

This analysis provides us with the two lessons regarding IBMs which
we will use, namely, the generic appearance of hidden
sector gauge groups with matter in the bifundamental representation,
and the generic appearance of $U(1)$ mixed anomalies which are
fixed by the Green-Schwarz mechanism.

\section{D-term Inflation}

Suppose we have $N$ brane stacks in the hidden sector.  From
the preceding discussion, we see that the diagonal subgroups
living on each stack will provide $N$ $D$-term equations, while
we expect to have ${\cal O}(N^2)$ scalars arising from the
strings stretching between these branes.  The first corollary
of our above general features is that there will be many
$D$-flat directions.  We plan to use this feature to provide
for a flat inflaton potential.

Basically, we can summarize the plan as follows
\begin{itemize}
\item{First, separate off a term $V_{\rm inf} ^{D}$ from
the remaining $D$-terms $V_{\rm rest} ^D$.  Schematically,
we have $V_{\rm inf} ^{D}= {g^2 \over 2}(|\phi_+|^2 -
|\phi_-|^2 -\xi)^2$, where $\phi_+$ is the waterfall
field which ends inflation.}
\item{Move out along a flat direction of $V^D$.}
\item{A generic Yukawa coupling of the form
$W=\lambda S \phi_+ \phi_-$ will generate a
potential for the waterfall field of the form
$V_F = \lambda^2 |S^2||\phi_+|^2 +...$.  If $S$ gets
a vev when moving out on the $D$-flat direction,
this mass will lift the waterfall field and allow inflation
to proceed as normal in $D$-term inflation.}
\item{A 1-loop Coleman-Weinberg potential of the
form $V=V_0 (1+{g^2 \over 4\pi^2}\log {\lambda^2 S^2 \over
\Lambda^2})$ will cause the inflaton $S$ to slowly roll
back.  When it reaches a critical value, the waterfall
field becomes tachyonic and inflation ends.}
\item{The extra gauge symmetry keeps the inflaton
direction flat by suppressing superpotential corrections.}
\end{itemize}
This plan follows the standard form for generating
$D$-term inflation\cite{Dterm}; we see how this plan can be
realized in the IBM setup\cite{Dutta:2007cr}.  One point to note is
that in the standard $D$-term inflation picture, one
must always make sure that the inflaton direction is
flat.  We have chosen a $D$-flat direction for the
inflaton, but one must be sure that superpotential
terms do not lift this flat direction (for example,
a term of the form $W=mS^2$ could lift one out of the
slow-roll regime).  In typical field theory models,
one usually relies on $R$-symmetry to ensure that
such tree-level mass terms are projected out.  One
cannot necessarily guarantee that the right discrete
symmetry exists in any specific model, though.  In our
IBM setup, we see that the extra gauge invariance of
the hidden sector saves us.  Gauge invariance prevents
the existence of a large tree-level mass for the
inflaton.  And as usual for $D$-term inflation, K\"ahler
corrections to the $F$-term potential are small because
$V_F \ll V_D$.

Furthermore, the $D$-flat direction which is used for
the inflaton must involve turning on more than one
field.  One can see this simply by noting that each
field is charged under two gauge groups, and thus
contributes to two $D$-terms.  Turning on such a field
alone cannot be a flat direction.  Instead, the fields
which are turned on must form an oriented polygon in
the associated quiver diagram (i.e., for each $U(1)$
factor we must turn on two fields which are oppositely
charged).  In this way, the positive contribution to
any $D$-term by one scalar field is compensated by the
negative contribution by the next scalar.

\begin{figure}
\includegraphics[width=10cm]{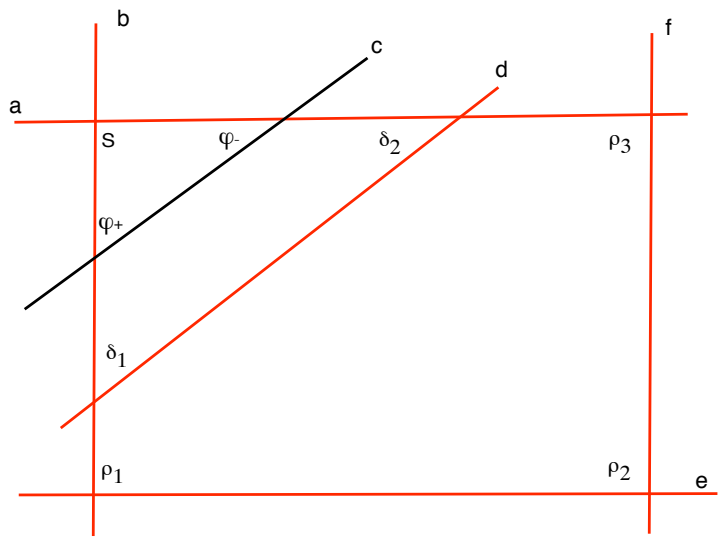}
\caption{This shows a setup of intersecting branes in type IIA that can yield $D$-term inflation.
We have a total of 6 branes. Brane $c$ is our inflationary brane.
Branes $a$ and $b$ give rises to the necessary Yukawa coupling between the
inflaton and the waterfall field.  Branes $e$ and $f$ are there to ensure that
the $D$-term potential has a flat direction.  This flat direction is broken by the
$F$-term potential but this is planckian suppressed. Finally,
brane $d$ is necessary for the anomalies to cancel, it plays no other role
during the inflationary era. $S$, $\phi_\pm$, $\delta_{1,2}$ and $\rho_{1,2,3}$ are the various
bifundamental fields at each intersections. Also note that this picture is a
two dimensional depiction of a configuration that actually lives in six curved dimensions.}
\label{fig:1}       
\end{figure}

The diagram in fig.\ \ref{fig:1} is illustrative.  This is
a schematic example of an IBM hidden sector which can
exhibit $D$-term inflation.  As will be clear,
the details of this model
are not essential to the construction; many other models
will work equally well.
In this
example, the brane $c$ is the inflationary brane, yielding
\begin{eqnarray}
V_{\rm{inf}} ^D &=& g_c ^2 (|\phi_+| ^2 -|\phi_-| ^2 -\xi_c)^2\; ,
\nonumber\\
V_{\rm{rest}} ^D &=&
g_a ^2 (|\phi_-| ^2 -|S| ^2 +|\rho_3|^2-\xi_a)^2
\nonumber\\
&\,& +g_b ^2 (|S| ^2 -|\phi_+| ^2 -|\rho_1|^2 -\xi_b)^2
\nonumber\\
&\,&+g_e ^2 (|\rho_1|^2 -|\rho_2|^2 -\xi_e)^2
\nonumber\\
&\,& +g_f ^2 (|\rho_2|^2 -|\rho_3|^2 -\xi_f)^2\, .
\nonumber
\end{eqnarray}
For simplicity, we have chosen $\xi_d=\delta_1=\delta_2=0$.
The waterfall field $\phi_+$ gets a tachyonic $mass^2$ from
$V_{\rm{inf}}$, and inflation ends when it
condenses.  The inflaton
is the flat direction which arises from turning on the
four fields $S$ and $\rho_{1,2,3}$ living at the corners of
the ``square."

A worldsheet instanton arises from a string worldsheet
stretching along the triangle bounded by the $a$, $b$
and $c$ branes.  This instanton generates the Yukawa
coupling $W=\lambda S \phi_1 \phi_2$.  When $S$ gets
a vev, the potential term
\begin{eqnarray}
V_F = \lambda^2 (|S|^2 |\phi_+|^2
+|S|^2 |\phi_-|^2 +|\phi_+|^2 |\phi_-|^2)
\nonumber
\end{eqnarray}
lifts the $mass^2$ of the waterfall field, making it non-tachyonic.
But the one-loop Coleman-Weinberg
potential $V=V_0 (1+{g^2 \over 4\pi^2}\log {\lambda^2 S^2 \over
\Lambda^2})$ causes $S$ to roll back along the $D$-flat direction
until the waterfall field becomes tachyonic, ending inflation.

The only Yukawa coupling which could lift the $D$-flat direction
is of the form
\begin{eqnarray}
W = {\lambda' \over M_p}  S \rho_1 \rho_2 \rho_3 ,
\nonumber
\end{eqnarray}
which is $M_p$-suppressed.
The polygon formed by the branes
whose intersections yield the inflaton flat direction could
have more than four sides.  A larger polygon would result in
more $M_p$ factors in the Yukawa coupling denominator, and
greater suppression of the $F$-terms which
might lift the inflaton direction.
For this simple ``square" example, as is often the case
in inflationary models, $M_p$ suppression is not necessarily
enough.  To stay in the slow-roll regime one must have
$g^2 \lambda^2 \leq 10^{-13}$.

Since $\lambda$ is generated
by a worldsheet instanton and is thus exponentially suppressed,
this tuning might not be unreasonable.  But in any case, this arises
only in this simple example; for a slightly more complicated
example where the flat direction comes from a ``pentagon" or
larger polygon, the fine-tuning is much less severe.
Thus we see one of the
advantages of this IBM setup; {\it the numerical fine-tuning of
the curvature of the potential is replaced by a discrete
fine-tuning of the sign of brane intersection numbers}.

The observational constraints from cosmology are
the correct size of density perturbations
($P_{\cal R} \sim 10^{-9}$) and the spectral index ($n_s\sim 1$),
and that there be at least 60 efolds of inflation.
In this setup, for moderate-sized Yukawa couplings one
finds the standard $D$-term inflation prediction of a
slightly red spectrum, $n_s \sim 0.98$.  This value is
not optimal, yet not inconsistent with the latest WMAP
data \cite{Spergel:2006hy}.  In this regime
one has $\xi \sim 10^{-5} M_p ^2$, which gives us a
cosmic string tension $G\mu \sim 10^{-5}$.  Observations
rule out any stable cosmic strings of tension
$G\mu > 10^{-7}$\cite{Wyman:2005tu}, however.
Fortunately, very simple modifications
of this setup can avoid this difficulty be ensuring that
the strings are unstable.  For example, one could have
$I_{ac} >1$, ensuring that there is more than one waterfall
field, or alternatively, gauge groups $U(N)$ with $N>1$. In
either case, $\Pi_1$ of the vacuum manifold is trivial, and
stable cosmic strings do not form.

Note that we have not discussed moduli stabilization in this
setup.  Instead, we have simply assumed that closed string
moduli are stabilized at a scale well above the scale of
inflation.  A variety of moduli stabilization schemes are
now known in Type II string theory.  It would be of great
interest to understand in detail how moduli can be stabilized
in a way consistent with this inflation setup, and in
particular with $V_F < V_D$.

\section{Hidden Sector Baryogenesis}
\label{sec:2}
We will see that the generic presence of mixed anomalies can
also affect cosmology.  One of the interesting questions about
cosmology is the origin of the observed baryon asymmetry.  The
process which generates this asymmetry is called baryogenesis,
and Sakharov showed that
the three conditions required for it to
occur are $B$ violation, $CP$ violation  and a departure from
thermal equilibrium\cite{Sakharov:1967dj}.

Consider a common scenario in intersecting brane models, wherein
the $SU(3)_{qcd}$ gauge group arises as a subgroup of a $U(3)$
gauge group living on a stack of D-branes.  The diagonal $U(1)_B$
subgroup has baryon number as a charge.  From our previous discussion,
we see that a generic feature of these intersecting brane models
is that this QCD brane stack will intersect with other hidden sector
branes, resulting in non-vectorlike matter charged under $U(1)_B$.
If $G$ is a gauge group living on a hidden sector brane with such
a topological intersection, the result is a $[U(1)_B G^2]$
mixed anomaly.  This leads to a non-trivial divergence in the
baryon current:\
\begin{eqnarray}
\partial_{\mu} J_B ^{\mu} \propto Tr [F_G \wedge F_G].
\nonumber
\end{eqnarray}

This non-vanishing divergence provides precisely the
the type of baryon number violation required for
baryogenesis\cite{Dutta:2006pt}.
In particular, sphaleron or instanton processes in the
hidden $G$ sector which transition between different vacua
will result in a shift in the baryon number.  In any
phenomenologically viable IBM, the gauge group $G$ must either
break or confine, in order to avoid the presence of massless exotic
fermions charged under $SU(3)_{qcd}$ and $G$.  If the associated
phase transition is first order (and accompanied by CP-violation,
which is generically possible), then all Sakharov conditions
are satisfied and a baryon asymmetry can be generated.

This mechanism of hidden sector baryogenesis seems very reminiscent
of electroweak baryogenesis\cite{Kuzmin:1985mm}, wherein
sphalerons in the electroweak
sector drive the baryon asymmetry.  EWBG is an interesting and
well-studied model, partly because it is very concrete and in
specific implementations (such as the SM or MSSM) can be analyzed
precisely and in detail.  In this regard, it is almost too successful;
EWBG is ruled out in the SM, and can only fit into a very narrow
window of MSSM parameter space ($m_h<120\, {\rm GeV}$,
$120\, {\rm GeV} <m_{stop}<m_{top}$).

What we see in this intersecting brane model context is that the
electroweak gauge group is just one group.  One expects several
other groups to appear which also have mixed anomalies with
$U(1)_B$, and for a realistic model all of these groups must exhibit
some phase transition.  Any of them can just as well generate a baryon
asymmetry through sphalerons at a symmetry breaking transition.
If even one of these hidden sector groups has a first order transition,
HSB can work.  In this sense, it is a quite robust feature of IBMs.

One might worry that these hidden sectors could break at a scale
higher than the electroweak scale, with electroweak sphalerons
washing out the generated asymmetry.  But in general IBMs, there is
a $U(1)_{B-L}$ anomaly as well as a $U(1)_B$ mixed anomaly.
Here, $U(1)_L$ is the gauge theory living on the brane where all leptonic
strings end.
A $[U(1)_L G^2]$ mixed anomaly will arise if this leptonic stack
has non-zero topological intersection with
the $G$ stack.  This is generically the case, and there is no reason for
$U(1)_B$ and $U(1)_L$ to have the same intersection numbers with $G$ (unless
those branes are parallel, as in a Pati-Salam model).
Since the $U(1)_B$ and $U(1)_L$ anomalies will then have different coefficients,
the $U(1)_{B-L}$ anomaly will not cancel.
Thus, the symmetry breaking transition
for $G$ can generally occur at any scale (including high
scales), and the $U(1)_{B-L}$ asymmetry ensures that the
generated baryon asymmetry cannot be washed out.

Interestingly, hidden sector baryogenesis can arise at the end of the
inflationary scenario described in section 3, with the inflationary gauge
sector acting as the hidden sector.  There will generically be chiral
matter charged under both the inflationary sector and $U(1)_B$, contributing
to a mixed anomaly which generates a non-trivial divergence of the
baryon current.
When the waterfall field
condenses to end inflation, energy is dumped into the hidden sector
through a process called tachyonic preheating\cite{Felder:2000hj}.  This
process will excite long-wavelength modes, including sphalerons
which drive baryon violation.
This fast process
necessarily occurs out of thermal equilibrium and can be accompanied
by CP-violation, thus satisfying the Sakharov conditions.
Thus, the scale of inflation and the scale of baryogenesis are tied
together.
This type
of baryogenesis was studied in the context of the electroweak group
\cite{EWTpreheat}.  In that context, however,
inflation would have to occur at the electroweak scale, which can
be difficult to reconcile with observation.  We see that in the
context of our inflationary scenario, a higher inflation scale is
possible.

\section{Conclusions}

One of the unique features of string theory is the ability to create
unified models of quantum gravity, matter and gauge theory.  As such,
one would hope that these models would provide insight useful to
both cosmology and phenomenology.
We have seen that intersecting brane models have several general features
which lead to interesting models of cosmology.  In a similar vein, one also
finds that these general string features have phenomenological applications
to dynamical supersymmetry breaking\cite{Kumar:2007dw} and other
signatures at the LHC\cite{stringLHC}.  Given the current and upcoming observational data
from cosmology and the prospect of new data coming soon from LHC, it
is worthwhile to study new physics scenarios which can provide insight
relevant to both types of data.

{\bf Acknowledgments} This work is supported in part by
NSF grant PHY-0314712 and PHY-0555575.

%

%

\begin{thebibliography}{999}
%
%

\bibitem{reviewinflation}
G.~R.~Dvali and S.~H.~H.~Tye,
Phys.\ Lett.\ B {\bf 450} (1999) 72;
  F.~Quevedo,
  Class.\ Quant.\ Grav.\  {\bf 19}, 5721 (2002);
  C.~P.~Burgess,
  arXiv:hep-th/0606020;
  S.~H.~Henry Tye,
  arXiv:hep-th/0610221;
  J.~M.~Cline,
  arXiv:hep-th/0612129;
  R.~Kallosh,
  arXiv:hep-th/0702059;
  E.~Silverstein and D.~Tong,
  Phys.\ Rev.\ D {\bf 70} (2004) 103505;
M.~Alishahiha, E.~Silverstein and D.~Tong,
  Phys.\ Rev.\ D {\bf 70} (2004) 123505;
  S.~Kachru, R.~Kallosh, A.~Linde, J.~M.~Maldacena, L.~McAllister and S.~P.~Trivedi,
  JCAP {\bf 0310}, 013 (2003).


\bibitem{IBM}
  R.~Blumenhagen, L.~Goerlich, B.~Kors and D.~Lust,
  JHEP {\bf 0010}, 006 (2000);
  C.~Angelantonj, I.~Antoniadis, E.~Dudas and A.~Sagnotti,
  Phys.\ Lett.\ B {\bf 489}, 223 (2000);
  R.~Blumenhagen, L.~Goerlich, B.~Kors and D.~Lust,
  Fortsch.\ Phys.\  {\bf 49}, 591 (2001);
  G.~Aldazabal, S.~Franco, L.~E.~Ibanez, R.~Rabadan and A.~M.~Uranga,
  J.\ Math.\ Phys.\  {\bf 42}, 3103 (2001);
  G.~Aldazabal, S.~Franco, L.~E.~Ibanez, R.~Rabadan and A.~M.~Uranga,
  JHEP {\bf 0102}, 047 (2001);
  M.~Cvetic, G.~Shiu and A.~M.~Uranga,
  Nucl.\ Phys.\ B {\bf 615}, 3 (2001);
  M.~Cvetic, P.~Langacker and G.~Shiu,
  Nucl.\ Phys.\ B {\bf 642}, 139 (2002);
  R.~Blumenhagen, V.~Braun, B.~Kors and D.~Lust,
  arXiv:hep-th/0210083;
  A.~M.~Uranga,
  Class.\ Quant.\ Grav.\  {\bf 20}, S373 (2003);
  M.~Cvetic and T.~Liu,
  Phys.\ Lett.\  B {\bf 610}, 122 (2005);
  F.~Marchesano and G.~Shiu,
  JHEP {\bf 0411}, 041 (2004).



\bibitem{Dterm}
  P.~Binetruy and G.~R.~Dvali,
  Phys.\ Lett.\  B {\bf 388}, 241 (1996);
  E.~Halyo,
  Phys.\ Lett.\  B {\bf 387}, 43 (1996).

\bibitem{Dutta:2007cr}
  B.~Dutta, J.~Kumar and L.~Leblond,
  JHEP {\bf 0707}, 045 (2007).

\bibitem{Spergel:2006hy}
  D.~N.~Spergel {\it et al.}  [WMAP Collaboration],
  arXiv:astro-ph/0603449.

\bibitem{Wyman:2005tu}
  M.~Wyman, L.~Pogosian and I.~Wasserman,
  Phys.\ Rev.\  D {\bf 72}, 023513 (2005)
  [Erratum-ibid.\  D {\bf 73}, 089905 (2006)].

\bibitem{Sakharov:1967dj}
  A.~D.~Sakharov,
  Pisma Zh.\ Eksp.\ Teor.\ Fiz.\  {\bf 5}, 32 (1967).

\bibitem{Dutta:2006pt}
  B.~Dutta and J.~Kumar,
  Phys.\ Lett.\  B {\bf 643}, 284 (2006).

\bibitem{Kuzmin:1985mm}
  V.~A.~Kuzmin, V.~A.~Rubakov and M.~E.~Shaposhnikov,
  Phys.\ Lett.\ B {\bf 155}, 36 (1985).

\bibitem{Felder:2000hj}
  G.~N.~Felder, J.~Garcia-Bellido, P.~B.~Greene, L.~Kofman, A.~D.~Linde and I.~Tkachev,
  Phys.\ Rev.\ Lett.\  {\bf 87}, 011601 (2001).

\bibitem{EWTpreheat}
  J.~Garcia-Bellido, M.~Garcia Perez and A.~Gonzalez-Arroyo,
  Phys.\ Rev.\  D {\bf 67}, 103501 (2003);
  J.~Smit and A.~Tranberg,
  JHEP {\bf 0212}, 020 (2002).

\bibitem{Kumar:2007dw}
  J.~Kumar,
  arXiv:0708.4116 [hep-th].

\bibitem{stringLHC}
for example, see
  P.~Anastasopoulos, M.~Bianchi, E.~Dudas and E.~Kiritsis,
  JHEP {\bf 0611}, 057 (2006);
  D.~Berenstein and S.~Pinansky,
  Phys.\ Rev.\  D {\bf 75}, 095009 (2007).









\end{thebibliography}
%

\end{document}